\documentclass[twocolumn,letter]{jpsj3}
\newcommand{\tc}{$T_{\rm c}$}
\newcommand{\msr}{$\mu$SR}

\def\runauthor{Kazuki \textsc{Ohishi} \textit{et al}.}

\title{No anomalous spin fluctuation in optimally doped and overdoped iron-based oxypnictide 
superconductors LaFeAsO$_{1-x}$F$_x$ probed by muon spin relaxation}
\author{
Kazuki \textsc{Ohishi}$^{1,5}$\thanks{E-mail:kazuki.ohishi@riken.jp}, 
Yasuyuki \textsc{Ishii}$^{1,5}$, Kiichi \textsc{Miyazawa}$^{2,3,5}$, 
Hideto \textsc{Fukazawa}$^{4,5}$, Isao \textsc{Watanabe}$^{1,5}$, 
Yoh \textsc{Kohori}$^{4,5}$, Parasharam M. \textsc{Shirage}$^{2,5}$, 
Hijiri \textsc{Kito}$^{2,5}$, Akira \textsc{Iyo}$^{2,3,5}$, Hiroshi \textsc{Eisaki}$^{2,5}$
}
\inst{
$^1$Advanced Meson Science Laboratory, Nishina Center for Accelerator-Based Science, 
RIKEN, Wako, Saitama 351-0198
\\
$^2$National Institute of Advanced Industrial Science and Technology (AIST), Tsukuba, Ibaraki 305-8568
\\
$^3$Department of Applied Electronics, Tokyo University of Science, Noda, Chiba 278-8510
\\
$^4$Department of Physics, Chiba University, Chiba 263-8522
\\
$^5$JST, Transformative Research-Project on Iron Pnictides (TRIP), Chiyoda-ku, Tokyo 102-0075
}
\abst{
Zero field and longitudinal field muon spin relaxation measurements have been performed in optimally doped and 
overdoped superconductors LaFeAsO$_{1-x}$F$_x$ in order to investigate the magnetic fluctuation 
over a wide range of temperature and longitudinal field. 
We have observed no sign of magnetic fluctuation against temperature in the muons' time window (10$^{-5}\sim$
10$^{-9}$~s). 
Considering the current results and other results, i.e., spin fluctuation observed by neutron scattering, pseudogap-like 
behaviors by NMR and photoemission spectroscopy, it is suggested that  
not only the spin fluctuation but also the multiband character with several different orbital contributions at the Fermi
surface may play an important role in the superconducting pairing mechanism of LaFeAsO$_{1-x}$F$_x$.}

\kword{superconductivity, normal state, iron-based oxypnictide, LaFeAsO$_{1-x}$F$_x$, \msr}

\begin{document}
\maketitle

Since the discovery of iron-based pnictide superconductor LaFeAsO$_{1-x}$F$_x$ (La1111) \cite{Kamihara2008} 
and related iron-based superconductors \cite{Kamihara2008,XHChen2008,Ren2008,GFChen2008,Kito2008,
Ren2008a,Sefat2008,Sefat2008a,Gao2008,Matsuishi2008,Rotter2008,Sasmal2008,GFChen2008a,GWHChen2008,
Miyazawa2009}, 
a tremendous number of studies have been stimulated. 
The parent compound LaFeAsO exhibits an antiferromagnetic 
(AF) ordered state. The AF phase abruptly vanishes with the electron doping 
by substitution of F for O \cite{Kamihara2008} or by oxygen deficiency \cite{Kito2008,Ren2008a,Miyazawa2009}, 
and the superconducting phase appears. 
As for the superconducting gap structure, it is argued that the AF spin fluctuations 
originate from the nesting between the two dimensional cylindrical Fermi surface (two hole surfaces around 
$\Gamma$ point and two electron surfaces around $M$ point), giving raise to the sign reversing $s_\pm$-wave 
superconducting state \cite{Mazin2008,Kuroki2008,Yanagi2008,Ikeda2008,Graser2009,Nomura2009,Wang2009}. 
Many experimental results observed in 1111-system, i.e., angle-resolved photoemission spectroscopy \cite{Liu2008}, 
NMR \cite{Nakai2008}, and muon spin relaxation (\msr\/) \cite{Takeshita2008}, support this multiple-isotropic gap 
scenario, while a multiple-nodal line gap scenario is proposed in some superconductors, 
i.e., LaFePO\cite{Fletcher2009,Kuroki2009,Yamashita2009} and KFe$_2$As$_2$ \cite{Fukazawa2009}. 

The spin lattice relaxation rate $1/T_1$ observed by NMR in the normal state revealed that the strong 
AF fluctuations in the parent compound are dramatically suppressed with electron doping, and a pseudogap-like behavior 
was observed in LaFeAsO$_{1-x}$F$_x$ ($x\ge 0.1$) \cite{Nakai2008,Grafe2008,Ahilan2008,Mukuda2009}. 
Such a pseudogap-like behavior has been also observed with the photoemissoin spectroscopy \cite{Sato2008}. 
It has been attracted much interests whether or not the origin of the pseudogap-like behavior is due to AF spin 
fluctuation. According to the recent theoretical approach, it is suggested that this phenomenon is not 
entirely of the magnetic origin but of specific band structure origin \cite{Ikeda2008}. 
This implies that the pairing mechanism cannot be attributed solely to the AF spin fluctuations. 
The \msr\/ technique is good tool to study the spin fluctuation because it has a unique time window compared 
with neutron scattering and NMR measurements. 
Regarding a lot of previous \msr\/ measurements in 1111-system \cite{Takeshita2008,Klauss2008,
Luetkens2008,Khasanov2008,Aczel2008,Carlo2009,Luetkens2009,Drew2009,Takeshita2009,Sanna2009,Maeter2009}, 
all reports has been focused on the magnetic ordered state and superconducting properties at the temperatures 
below transition temperatures ($T_{\rm SDW}$ or \tc\/), so that there is no report studying the magnetic 
fluctuation far above \tc\/, namely, in the normal state. 
In order to investigate the magnetic property microscopically over wide temperature range in La1111 system, 
we have performed zero field (ZF) and longitudinal field (LF) \msr\/ measurements in optimally doped and overdoped 
high quality La1111 samples. 

\begin{table*}[t]
\caption{\tc\/, lattice parameters at room temperature, rms width of fluctuating field and correlation time 
determined by current measurements in LaFeAsO$_{1-x}$F$_x$ samples. }
\label{t1}
\begin{tabular}{lccccc}\hline
\multicolumn{1}{c}{Samples} & \multicolumn{1}{c}{\tc\/ [K]} & \multicolumn{1}{c}{$a$ [\AA]} & \multicolumn{1}{c}{$c$ [\AA]} & \multicolumn{1}{c}{$\mu_0\delta B$ [$\mu$T]}  & \multicolumn{1}{c}{$\tau_{\rm c}$ [$\mu$s]} \\
\hline
La1111-27K (optimally doped) & 27 & 4.0264 & 8.7170 & 41(7) & 0.8(1) \\
La1111-23K (overdoped) & 23 & 4.0248 & 8.7160 & 31(2) & 2.6(4) \\
La1111-19K (overdoped) & 19 & 4.0244 & 8.7140 & 58(4) & 0.8(3) \\
\hline
\end{tabular}
\end{table*}

Polycrystalline samples were synthesized by solid state reaction. The detailed procedure for sample preparation is 
described in ref.~\citen{Miyazawa2008}. 
The samples were confirmed to be single phase using X-ray diffraction analysis. In addition, we confirmed that 
no ferromagnetic impurity exists in our samples by magnetization measurements. The values of \tc\/ were 
determined by susceptibility measurements, which exhibit marked decrease due to 
the onset of superconducting diamagnetism. 
Considering the relation between \tc\/ and lattice parameters, we have confirmed 
that prepared samples were optimally doped and overdoped samples with 
\tc\/ = 27, 23, and 19~K, respectively, as summarized in Table~\ref{t1}. 

Conventional time-differential \msr\/ experiments were carried out 
at the RIKEN-RAL Muon Facility in the UK which provides pulsed beam of nearly 100\% spin polarized muons. 
The samples were mounted on a sample holder made of 99.995\%\/ silver and loaded to the appropriate cryostat.
A fly-past setup was used to extremely reduce the background signals from muons which missed the sample. 
ZF-\msr\/ measurements were performed at temperatures between 4~K and 
room temperature. The dynamics of local magnetic fields at muon sites were investigated by \msr\/ measurements under a LF \cite{Schenck1986}. 

\begin{figure}
\begin{center}
\includegraphics[width=0.4\textwidth]{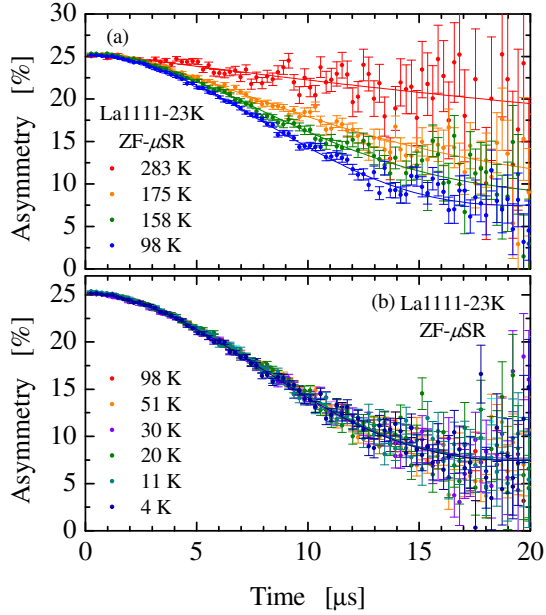}
\end{center}
\caption{(Color online) ZF-\msr\/ time spectra in La1111-23K at various temperatures.}
\label{ZF23K} 
\end{figure}
\begin{figure}
\begin{center}
\includegraphics[width=0.4\textwidth]{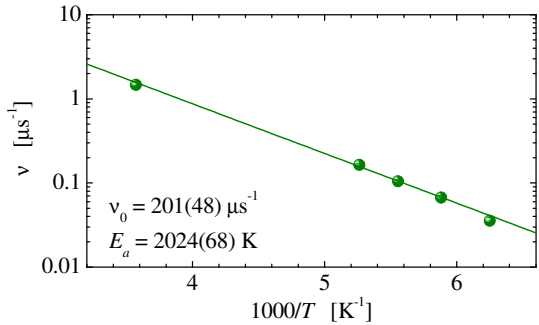}
\end{center}
\caption{(Color online) Muon hopping rate $\nu$ against the inverse of temperature. 
The solid line is fit to the model given by eq.~(\ref{thermal}).}
\label{nu} 
\end{figure}

\begin{figure}
\begin{center}
\includegraphics[width=0.4\textwidth]{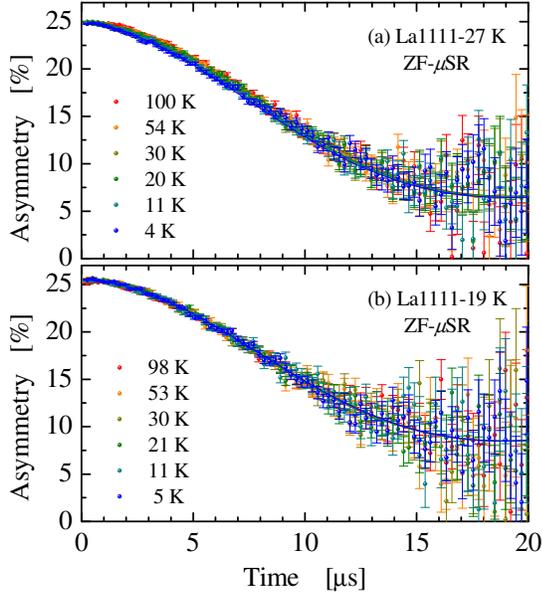}
\end{center}
\caption{(Color online) ZF-\msr\/ time spectra in (a) La1111-27K and 
(b) La1111-19K at various temperatures. }
\label{ZF27and19K} 
\end{figure}

\begin{figure}
\begin{center}
\includegraphics[width=0.4\textwidth]{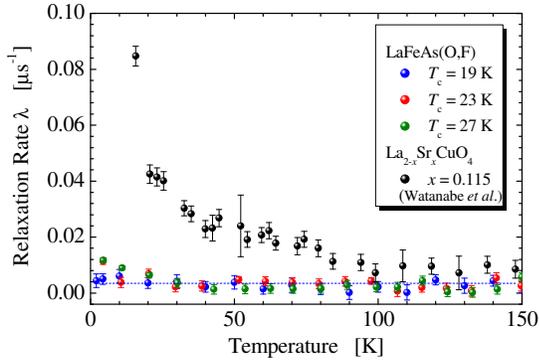}
\end{center}
\caption{(Color online) Temperature dependence of the relaxation rate $\lambda$ in 
La1111-27K, La1111-23K and La1111-19K from the present work as well as that in 
La$_{2-x}$Sr$_x$CuO$_4$ ($x=0.115$) \cite{Watanabe2008}. }
\label{ZF-lambda} 
\end{figure}

ZF-\msr\/ time spectra obtained in all samples above 150~K were well fitted 
by a dynamic Kubo-Toyabe (DKT) function \cite{Hayano1979}, 
\begin{equation}
AP_z(t)=A_{\rm s}G_{\rm DKT}(\Delta,\nu,t)+A_{\rm Ag},
\label{DKT}
\end{equation}
where $A_{\rm s}$ and $A_{\rm Ag}$ are the partial $\mu$-$e$ decay asymmetries ($A=A_{\rm s}+A_{\rm Ag}$) 
for sample and the sample holder, respectively, $P_z(t)$ is the time evolution of muon spin polarization, 
$G_{\rm DKT}(\Delta,\nu,t)$ is the DKT function, 
$\Delta$ is the second moment of the local magnetic field due to the nuclear moments, 
$\nu$ is the muon hopping rate. 
Figure~\ref{ZF23K}(a) shows the ZF-\msr\/ time spectra observed in 
the sample of La1111 with \tc\/~=~23~K (La1111-23K) obtained above $\sim$100~K. 
The solid curves in Fig.~\ref{ZF23K}(a) are the best fit results with eq.~(\ref{DKT}) with temperature 
independent values of $A_{\rm s}=18.3(2)$\%\/, $A_{\rm Ag}=6.8(2)$\%\/, $\Delta$ = 0.0901(7)~$\mu$s$^{-1}$ 
and temperature dependent parameter of $\nu$. 
The temperature dependence of $\nu$ observed above 150~K is well represented by the Arrhenius formula, 
\begin{equation}
\nu = \nu_0\exp\left(-\frac{E_a}{k_{\rm B}T}\right),
\label{thermal}
\end{equation}
where $E_a$ is the activation energy and $k_{\rm B}$ is the Boltzmann constant. 
As shown in Fig.~\ref{nu}, a fitting analysis with eq.~(\ref{nu}) yields 
$\nu_0 =2.01\pm0.48\times10^8$~s$^{-1}$ and $E_a =2024(68)$~K. 
This can be described in terms of a motional narrowing effect due to muons' motion in the sample 
at high temperatures above $\sim$150~K. 
Below 150~K, no change was observed in time evolution of muon spin polarization as shown in Fig.~\ref{ZF23K}(b). 
In order to clarify the absence of change in muon spin relaxation, these data were fitted by following function 
with the same $\Delta$,
\begin{equation}
AP_z(t)=A_{\rm s}\exp(-\lambda t)G_{\rm KT}(\Delta,t)+A_{\rm Ag}, 
\label{KT}
\end{equation}
where $\lambda$ is the relaxation rate, $G_{\rm KT}(\Delta,t)$ is a static Kubo-Toyabe function \cite{Hayano1979}. 
A fitting procedure similar to that for La1111-23K has been applied to the data observed 
in La1111-27K and La1111-19K. 
As shown in Fig.~\ref{ZF27and19K}, no change was observed in both samples, 
while the motional narrowing effect was also observed at high temperature (data are not shown). 
The values of $\Delta$ for La1111-27K and La1111-19K 
were estimated to be 0.0912(7) $\mu$s$^{-1}$ and 0.0914(8) $\mu$s$^{-1}$, respectively. 
Similar values of $\Delta$ were obtained in these three samples, indicating that muons basically probed the similar 
internal fields at the same position in all samples. 
Temperature dependence of $\lambda$ observed in all samples are shown in Fig.~\ref{ZF-lambda}. 
For comparison, $\lambda$ observed in high-\tc\/ cuprate 
La$_{2-x}$Sr$_x$CuO$_4$ ($x=0.115$) \cite{Watanabe2008} (LSCO) are also plotted. 
It is apparent that the dynamic depolarization rate $\lambda$ observed in all of La1111 samples are completely 
temperature independent, suggesting that 
there is no magnetic fluctuation in these materials. 

\begin{figure}
\begin{center}
\includegraphics[width=0.4\textwidth]{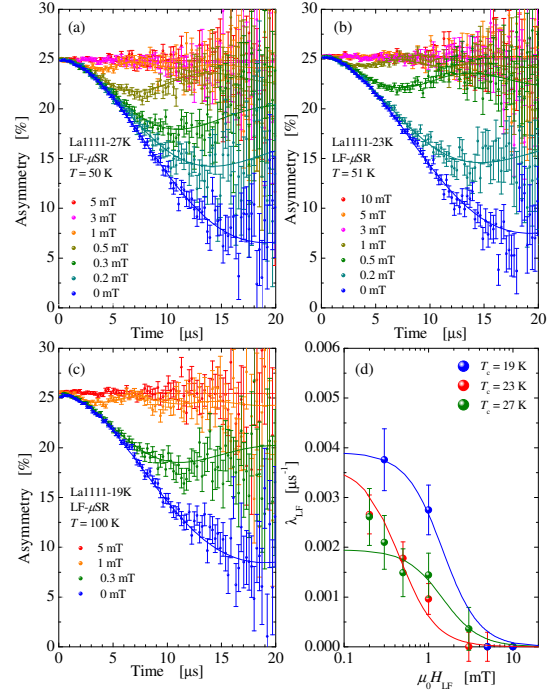}
\end{center}
\caption{(Color online) ZF- and LF-\msr\/ time spectra in (a) La1111-27K, (b) La1111-23K and (c) La1111-19K 
at 50 K, 51 K and 100 K, respectively. (d) Applied field dependence of the relaxation rate $\lambda_{\rm LF}$ in 
La1111-27K, La1111-23K and La1111-19K. The solid curves are fits to the model given by eq.~(\ref{lmd}). }
\label{LF}
\end{figure}

The LF-\msr\/ measurements were performed to evaluate the effect of spin dynamics and the fluctuating field 
in the normal state. 
Figure~\ref{LF} shows the LF-\msr\/ time spectra in all La1111 samples at various fields. 
The Gaussian-type time spectra observed under ZF are recovered with increasing a longitudinal 
magnetic field $H_{\rm LF}$ and fully decoupled with $\mu_0H_{\rm LF}=5$~mT. Here, eq.~(\ref{KT}) can 
be expanded for the case under a $H_{\rm LF}$ as 
\begin{equation}
AP_z(t)=A_{\rm s}\exp(-\lambda_{\rm LF} t)G_{\rm KT}(\Delta,H_{\rm LF},t)+A_{\rm Ag},
\label{LFKT}
\end{equation}
where $G_{\rm KT}(\Delta,H_{\rm LF},t)$ is the static Kubo-Toyabe function including the effect of $H_{\rm LF}$, and 
$A_{\rm s}$, $A_{\rm Ag}$ and $\Delta$ are fixed to the values obtained by analyzing ZF-\msr\/ spectra. 
The solid curves in Fig.~\ref{LF} (a)-(c) shows the fitting analysis by eq.~(\ref{LFKT}). 
The $H_{\rm LF}$ dependence of $\lambda_{\rm LF}$ shown in Fig.~\ref{LF} (d)  
can be fitted by the Redfield model \cite{Schenck1986}, 
\begin{equation}
\lambda_{\rm LF}=\frac{2\gamma_\mu^2(\mu_0\delta B)^2\tau_{\rm c}}{1+\gamma_\mu^2(\mu_0H_{\rm LF})^2\tau_{\rm c}^2},
\label{lmd}
\end{equation}
where $\gamma_\mu$ it the muon gyromagnetic ratio ($=2\pi\times 135.53$ MHz/T), $\delta B$ is the rms width of 
the fluctuating field, and $\tau_{\rm c}$ is the correlation time. The values of $\delta B$ and $\tau_{\rm c}$ for 
La1111-27K, La1111-23K and La1111-19K are obtained to be 41(7) $\mu$T and 0.8(1) $\mu$s, 31(2) $\mu$T 
and 2.6(4) $\mu$s and 58(4) $\mu$T and 0.8(3) $\mu$s, respectively. These values are summarized in Table~\ref{t1}. 

We have observed two different internal fields in La1111 samples. One of them is due to nuclear dipole fields 
expressed as $\Delta$, the other is due to spin fluctuation expressed as $\lambda$. 
Here, we discuss the latter parameter. 
As mentioned above, no temperature dependence of $\lambda$ has been observed around at both \tc\/ and 
the pseudogap-like energy about $\sim$170~K \cite{Nakai2008}, while $\lambda$ observed in 
LSCO starts to increase with decreasing temperature at around 100~K \cite{Watanabe2008}, suggesting that 
the spin fluctuation in LSCO is slow enough to be observed by muons (as mentioned later). 
The estimated values $\mu_0\delta B$ are quite small like several tens of $\mu$T, 
which are same as earth's magnetism, and their fluctuations are quite static. 
The observed values of $\mu_0\delta B$ are inconsistent with the previous \msr\/ results 
observed in underdoped La1111 \cite{Takeshita2008}, 
where there is a phase separation into superconducting and spin-glass-like magnetic phases. 
The estimated value of $\mu_0\delta B$ in ref.~\citen{Takeshita2008} is two order of larger than those 
in our results, suggesting that the origin of these fields are different, namely, that in underdoped sample 
can be a part of spin density wave.
Here, we have two possibilities to describe the observed static fluctuation. 
One is a hyperfine coupling between the electronic spins and muons, and the other is that between the electronic 
spins via nuclear spins and muons. Empirically, it is unlikely to suppose the former case because 
the observed fluctuation and internal field are extremely small compared with those due to the electronic spins, i.e., 
it is expected to be $\mu_0\delta B\sim$ mT \cite{Takeshita2008}. Hence, we find little compelling 
evidence for the former case. 

We now discuss possible case, in which it is too fast for muons to observe the spin fluctuation in the muons' 
time window (10$^{-5}\sim$10$^{-9}$~s). 
According to the recent neutron results of magnetic excitation in optimally doped La1111 
(\tc\/ = 25, 29~K) \cite{Wakimoto2009}, magnetic fluctuation was observed both above and below \tc\/, 
while it disappears in heavily-overdoped sample which does not show superconductivity. 
It concludes that the electronic spin fluctuations in optimally and overdoped samples are very fast compared with 
muons' time window so that we cannot observe such fluctuations by means of \msr\/. 
Needless to say, this explanation is consistent with the pseudogap-like behavior observed by NMR \cite{Nakai2008}, 
photoemission spectroscopy \cite{Sato2008} and also with the theoretical predictions 
\cite{Ikeda2008,Ikeda2010,Ikeda2010a}, in which it is insisted that the pseudogap-like behavior originates from 
the band structure effect. 
Especially, according to the recent calculation \cite{Ikeda2010a}, it is suggested that the low-energy spin excitation 
in electron doped is quite small compared with that in hole doped. 
It is obvious proof of this theoretical prediction that we observed no temperature dependence of $\lambda$. 
Considering the effect of spin fluctuation observed by \msr\/ and neutron scattering \cite{Wakimoto2009} and 
the pseudogap-like behaviors observed by NMR \cite{Nakai2008} and photoemission spectroscopy \cite{Sato2008}, 
not only the spin fluctuation but also the multiband character with several different orbital contributions at the Fermi
surface may play an important role in the superconducting mechanism of La1111, as it is suggested by the 
theoretical approach \cite{Ikeda2010}.

In summary, using magnetically sensitive \msr\/ technique we have revealed the no anomaly of spin fluctuation 
over a wide range of temperature in optimally doped and overdoped La1111 samples. 
The observed values of $\Delta = 0.090$~$\mu$s$^{-1}$ in all samples were consistent with each other. 
No temperature dependence of $\lambda$ was observed, while the motional 
narrowing effect due to muons' motion was observed at high temperature. 
Considering the current results and other experimental and theoretical results, it is suggested that 
both the spin fluctuation and the multiband character 
may play an important role in the superconducting pairing mechanism of LaFeAsO$_{1-x}$F$_x$.

\section*{Acknowledgments}
This work was partially supported by a Grant-in-Aid for Scientific Research from the Ministry of Education, Culture, Sports, Science and Technology (MEXT).

\end{document}